\documentclass[aps,prl,preprint,groupedaddress]{revtex4-1}

\usepackage{graphics}
\usepackage{graphicx}
\usepackage{epsfig}
\begin{document}

\title{Microscopic Description of the Odd-Even Effect in Cold Fission}


\author{M. Mirea}

\affiliation{Horia Hulubei National Institute for Physics and Nuclear Engineering,
P.O. Box MG-G, Bucharest, Romania}

\begin{abstract}
The time dependent equations of motion for the  pair breaking
effect were corroborated with a condition that fixes dynamically the number of particles
on the two fission fragment. The single particle level scheme was calculated
with the Woods-Saxon superasymmetric two center shell model. This model
provides a continuous variation of the energies from one nucleus up to two
separated fragments. The dissipated energy resorts
from the time dependent pairing equations. A peculiar phenomenon
was observed experimentally in cold fission: the odd partition yields are favored over the even ones.
This odd-even effect for
cold fission was explained microscopically.
\end{abstract}



\maketitle

\section{Introduction}
\label{introd}
By identifying unambiguously the fission fragments according to their
mass number and their charge, the fission yields as function of the excitation
energy were measured \cite{schwab}. The experimental data showed a dominance of 
odd-odd fragments at excitation energies close to zero for U isotopes. This phenomenon
was independently remarked also in Ref. \cite{ham,ham2} for Cf. 
A first interpretation involved
a proportionality between the level densities of the fission fragments
and the yields. 

Recently, a new set of time-dependent coupled channel equations
derived from the variational principle was proposed
to determine dynamically the mixing 
between seniority-zero and seniority-two configurations \cite{plb09}. 
The essential idea is that the
configuration mixing is managed under the action of some inherent
low lying time dependent excitations produced in the avoided crossing
regions, that is, a dynamical mechanism like the Landau-Zener effect \cite{hill}.

In the pioneering investigations of the mass and charge distributions at low
excitation energies, the experimental results
showed a preference for the mass division
leading to even-$Z$ fragments \cite{reis,wo}. It was believed that at very
low excitation energies, i.e., high kinetic energies, the fragments
will be fully paired. Improving the experimental
procedure, these facts were contradicted
in Ref. \cite{sign} where no strong even-odd effect was evidenced in
the thermal neutron induced fission for high values of the total
kinetic energy. Even and odd partitions were observed experimentally close
to their respective $Q$-values for four systems investigated:
$^{233,235}$U(n$_{th}$,f), $^{239}$Pu(n$_{th}$,f) and $^{252}$Cf(sf).
It is worth to underline that the importance of the Landau-Zener effect in
the cold fission fragmentation behavior was anticipated, as mentioned
in Ref. \cite{gon}.
However, the odd-even structure
structure in fission is explained usually within statistical
arguments, as for example in Refs. \cite{rejmund,avrig,montoya,cle}.
Recently, some arguments  linked the odd-even structure also to the charge asymmetry evolution
during the fission process \cite{skh,skh2},
not only to the dissipated energy  as has been done earlier \cite{arm,ni}.

\section{Formalism}

In this section, the ingredients required to investigate dynamically
the odd-even effect are described.

\subsection{The equations of motion}

The usual theories of fission consider that the nuclear system
is characterized by several generalized coordinates $q(t)=\{q_n(t)\}$, $(n=1,...,N)$.
These coordinates vary and force the system to split into two separated fragments. 
The single-particle energies, and the many-body wave function are determined
by the the variation in time of these coordinates.
In order to deduce the microscopic equations of motion, the starting point
is  a many-body Hamiltonian with pairing residual
interactions:
\begin{equation}
H(q_{i}(t))=\sum_{k>0} \epsilon_{k}(q_{i}(t))(a_{k}^{+}a_{k}+a_{\bar k}^{+}a_{\bar k})-\sum_{k,l>0}G_{kl}(t)a_{k}^{+}a_{\bar{k}}^{+}
a_{l}a_{\bar{l}}.
\label{hami1}
\end{equation}
This Hamiltonian depends on the collective parameters
$q(t)$, allowed to vary with respect the time.
Here, $\epsilon_k$ are single particle energies, 
$a_{k}^{+}$ and $a_{k}$ denote operators for creating
and destroying a particle in the state $k$, respectively. The state characterized
by a bar signifies the time-reversed partner of a pair.
The pairing correlations arise from the short range interaction of
correlated pairs of fermions moving in time-reversed orbits. 
$G_{kl}$ is the matrix element of the pairing interaction
and its value is in principle dependent on the overlap
of the wave functions of the pairs.
For a given nucleus, it is possible to
approximate the pairing interaction matrix elements with a constant value 
by using a renormalization procedure that depends on the number of states
in the active pairing space and on the structure \cite{nix}.

\begin{figure}
\resizebox{0.90\textwidth}{!}{
  \includegraphics{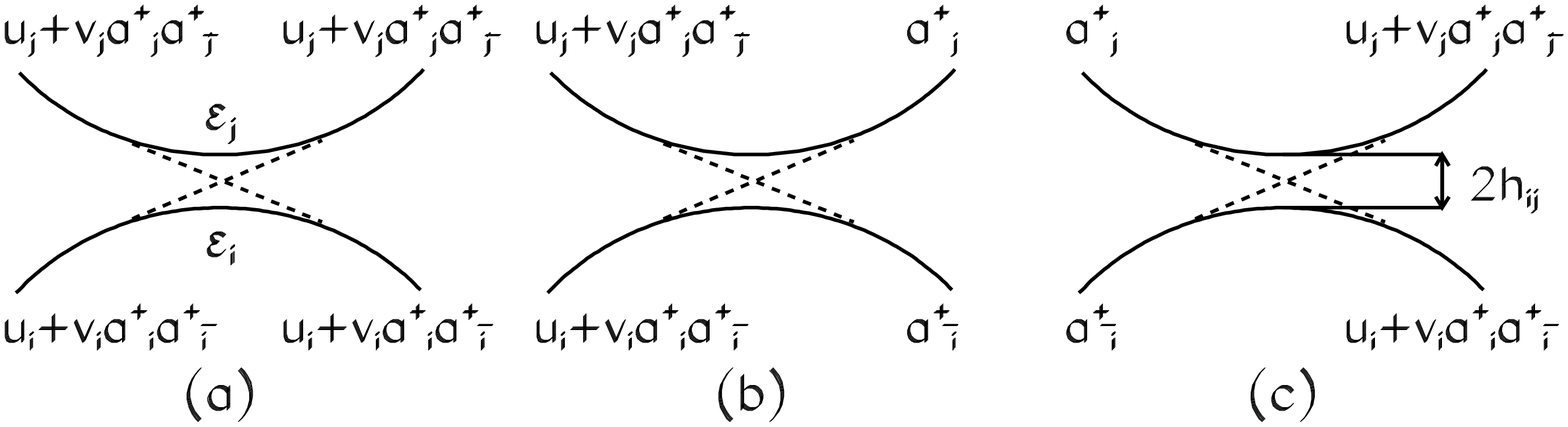}}
\caption
{
 Ideal avoided crossing regions between two adiabatic single
particle levels 
$\epsilon_i$ and $\epsilon_j$ characterized by the same
good quantum numbers. Three
possible transitions between configurations
in an avoided crossing region in the superfluid model are
displayed: (a) The configuration
remains unchanged after the passage through the avoided crossing region;
(b) A pair is broken; (c) A pair is created.
}
\label{figura1}
\end{figure}

Concerning the pair breaking effect, it must be evidenced that
only two types of velocity dependent excitation mechanisms
 are possible between different
single particle (or molecular) 
states of a dynamical nuclear system \cite{park}: the radial coupling
that can be described by the Landau-Zener effect in avoided levels
crossings regions and the Coriolis one produced in the region
of real crossings. The Coriolis coupling is responsible for transitions
between levels characterized by quantum numbers of the projection of
the spin that differ by one unity. This coupling is important in the treatment of
the $\alpha$-decay where the inertia is small and such an investigation
was made  in Ref. \cite{m01}.  Concerning the dynamical pair breaking,
it can be
described by a mechanism similar
to the Landau-Zener one.
The perturbation that produces the pair breaking in avoided
crossing regions between levels with the
same good quantum numbers is obtained in terms of
quasiparticle creation and annihilation operators 
\begin{equation}
\begin{array}{c}
\alpha_{k(ij)}=u_{k(ij)}a_{k}-v_{k(ij)}a_{\bar k}^{+};~~
\alpha_{\bar{k}(ij)}=u_{k(ij)}a_{\bar k}+v_{k(ij)}a_{ k}^{+};\\
\alpha_{k(ij)}^{+}=u_{k(ij)}a_{k}^{+}-v_{k(ij)}^{*}a_{\bar k};~~
\alpha_{\bar{k}(ij)}^{+}=u_{k(ij)}a_{\bar k}^{+}+v_{k(ij)}^{*}a_{ k}.
\label{anho2}
\end{array}
\end{equation}
Within the previous transformations, some perturbations that  
break dynamically a Cooper
pair when the system traverses an avoided level crossing region are constructed.
The parameters $v_{k(ij)}$ and $u_{k(ij)}$ are occupation and vacancy amplitudes,
respectively, for a pair occupying the single-particle level $k$ of the configuration $\{ij\}$.
The seniority-zero configuration is denoted with 0 and the
seniority-two configuration by a pair of indexes $\{ij\}$, $i$ and $j$ denoting the
levels occupied by unpaired fermions.
As evidenced in Ref. \cite{plb09}, the three situations plotted in Fig. \ref{figura1} can be
modeled. In the plot \ref{figura1}(a), the Cooper pair remains
on the adiabatic level $\epsilon_{j}$,  in \ref{figura1} (b) the pair destruction is
illustrated,
 while in \ref{figura1} (c) two fermions generate a pair. To describe
these three situations, a residual perturbation in the avoided level crossing region 
is postulated
as follows:
\begin{equation}
\begin{array}{ll}
H'(t)=\sum_{i,j\ne i}^{n}h_{ij}(t)   &
\left[\alpha_{i(0)}\alpha_{\bar{j}(0)}
\prod_{k\ne i,j}\alpha_{k(0)}\alpha_{k(ij)}^{+}\right.\\ \nonumber
 &  \left. +\alpha_{i(0)}^{+}\alpha_{\bar{j}(0)}^{+}
\prod_{k\ne i,j}\alpha_{k(ij)}\alpha_{k(0)}^{+}\right]
,
\end{array}
\label{cort}
\end{equation}
where $h_{ij}$ is the interaction energy between levels in the
avoided level crossing regions.  Under the action of the interaction
$h_{ij}$, according to the identities (\ref{proper}) of the Appendix,
the operators $\alpha_{i(0)}\alpha_{\bar{j}(0)}$ and $\alpha_{i(0)}^{+}\alpha_{\bar{j}(0)}^{+}$
transform a Bogoliubov 
seniority zero wave function into a seniority two one, and vice versa, being
responsible for configuration mixing. 
The products over the index $k\ne i,j$ in the previous formula transform
the remaining Bogoliubov amplitudes from values
pertaining to the seniority zero wave function to those
pertaining to the seniority two functions with unpaired orbitals
$i$ and $j$, and vice-versa.  This dynamical pair breaking effect was theoretically
formulated in Ref. \cite{plb09} for the first time. It is important to note that
the same kind of perturbation
was used also to generalize the Landau-Zener effect in superfluid systems \cite{prc08,mpla03,int}. It was
demonstrated in Ref. \cite{prc08} that equations governing the Landau-Zener effect 
and the time dependent pairing equations
are two particular cases of a new set of coupled channel equations.

The two fission products must be characterized by integer numbers
of neutrons and protons. As a consequence, the sums of the occupation
probabilities of single particle levels of the two fragments
must give the mass and charge numbers. 
 By solving the equations of motion, unfortunately, 
the sum of single-particle densities (BCS occupation probabilities) of the
single particle levels belonging to the two fission fragments obtained
after the scission don't give exactly their numbers of nucleons.
A recipe can be implemented to fix dynamically these numbers
of particles in the two final fragments by using the
operators for the number of particles
$\hat{N}_i$ $(i=1,2)$ that act on each fission product. 
At scission, the two fission fragments must be characterized
by a supplementary condition
\begin{equation}
\mid N_2\hat{N}_1-N_1\hat{N}_2\mid =0,
\label{coop}
\end{equation}
where $N_1$ and $N_2$ are the number of particles in the final
fragments labeled 1 and 2, respectively, and
\begin{equation}
\hat{N_1}=\sum_{k{1}}(a_{k_{1}}^{+}a_{k_{1}}+a_{\bar k_{1}}^{+}a_{\bar k_{1}}); ~~
\hat{N_2}=\sum_{k{2}}(a_{k_{2}}^{+}a_{k_{2}}+a_{\bar k_{2}}^{+}a_{\bar k_{2}})
\label{co}
\end{equation}
are the corresponding operators.
Here, $k_1$ and $k_2$ run over the pairing active level states that are
located in the potential wells of
 the final fragment 1 and of the final fragment 2, respectively. The condition
(\ref{coop}) can be introduced in the equations of motion by means of
the Lagrange multipliers \cite{prc11,plb12}.

All the previous ingredients could be used to obtain the microscopic equations
of motion.
These equations are obtained from the variational principle
by minimizing the following energy functional
\begin{equation}
\delta {\cal{L}}=\delta\left\langle\varphi\left\vert H+H'-
\lambda\mid N_2\hat{N}_{1}-N_1\hat{N}_{2}\mid-
i\hbar{\partial\over\partial t}\right\vert\varphi\right\rangle.
\label{var}
\end{equation}
The trial many-body function $\varphi$ is a superposition of Bogoliubov seniority zero and 
seniority two wave functions
\begin{equation} \begin{array}{ll}
\mid\varphi(t)\rangle=&\left[c_{0}\prod_{k}
\left(u_{k(0)}(t)+v_{k(0)}(t)a_{k}^{+}a_{\bar{k}}^{+}\right)\right.\\ \nonumber
 & \left.  +\sum_{j,l\ne j}c_{jl}(t)a_{j}^{+}a_{\bar{l}}^{+}\prod_{k\ne j,l}
\left(u_{k(jl)}(t)+v_{k(jl)}(t)a_{k}^{+}a_{\bar{k}}^{+}\right)\right]\mid 0\rangle.
\end{array}
\label{wf2}
\end{equation}
$c_0$ is the amplitude of the seniority zero wave function
while $c_{jl}$ are amplitudes for the seniority two wave functions
for configurations in which the single particle orbitals $j$ and $l$ belonging 
to the active pairing space are each blocked by only one unpaired nucleon.
Here, the vacancy $u_k$ and occupation $v_k$ amplitudes 
are not the adiabatic solutions of the BCS equations
and depend on the variation in time of the generalized parameters
and the history of the nuclear system. $\lambda$ is a Lagrange multiplier.

The evolution in time of the nuclear system, if the collective
parameters $\{q_n(t)\}$ ($n=1,...,N$) vary, is obtained by
 performing the variation of the functional (\ref{var}). 
The procedure required for the functional variation is described in details in
the Appendix.
The next coupled channel equations are obtained, eventually:

\begin{equation}
\label{ec1}
i\hbar \dot{\rho}_{k(0)}=\kappa_{k(0)}\Delta_{k(0)}^{*}-
\kappa_{k(0)}^{*}\Delta_{k(0)},
\end{equation}
\begin{equation}
\label{ec11}
i\hbar \dot{\rho}_{k(jl)}=\kappa_{k(jl)}\Delta_{k(jl)}^{*}-
\kappa_{k(jl)}^{*}\Delta_{k(jl)},
\end{equation}

\begin{equation}
\begin{array}{c}
i\hbar \dot{\kappa}_{k(0)}=\left(2\rho_{k(0)}-1\right)\Delta_{k(0)}+
2\kappa_{k(0)}\left(\epsilon_{k}-sN_{i_k}\lambda\right)\\
-2G_{kk}\rho_{k(0)}\kappa_{k(0)},
\end{array}
\label{ec22}
\end{equation}
\begin{equation}
\begin{array}{c}
i\hbar \dot{\kappa}_{k(jl)}=\left(2\rho_{k(jl)}-1\right)\Delta_{k(jl)}+
2\kappa_{k(jl)}\left(\epsilon_{k}-sN_{i_k}\lambda\right)\\
-2G_{kk}\rho_{k(jl)}\kappa_{k(jl)},
\end{array}
\label{ec2}
\end{equation}

\begin{equation}
i\hbar \dot{P}_{0}=\sum_{l,j\ne l}h_{lj}(S_{0jl}^{*}-S_{0jl})
\label{ec3}
\end{equation}

\begin{equation}
i\hbar \dot{P}_{jl}=h_{lj}(S_{0jl}-S_{0jl}^{*})
\label{ec4}
\end{equation}

\begin{equation}
\begin{array}{c}
i\hbar \dot{S}_{0jl}=S_{0jl}(\bar{E}_{0}-\bar{E}_{jl})+
S_{0jl}\left(\sum_{k\ne j,l}T_{k(jl)}-\sum_{k}T_{k(0)}\right)\\
+\sum_{\{mn\}\ne \{jl\}}h_{mn}S_{mnjl}+h_{jl}(P_{jl}-P_{0})
\end{array}
\label{ec5}
\end{equation}

\begin{equation}
\begin{array}{c}
i\hbar \dot{S}_{mnjl}=S_{mnjl}(\bar{E}_{mn}-\bar{E}_{jl})+
S_{mnjl}\left(\sum_{k\ne m,n}T_{k(mn)}-\sum_{k\ne j,l}T_{k(jl)}\right)\\
+h_{mn}S_{0jl}-h_{jl}S_{0mn}^{*}
\end{array}
\label{ec6}
\end{equation}
where $j$, $k$, $l$, $m$, $n$ label the single particle levels in the
active pairing space. The sign $s=\pm 1$ ensures that the matrix element of
the expression (\ref{coop}) is positive.  
$N_{i_k}=N_2$ or $N_{i_k}=-N_1$ if the state $k$ will be located in the fragment 1 
or in the fragment 2 after the scission, respectively.
Here, the following notations are used:
\begin{equation}
\begin{array}{c}
\Delta_{k(0)}=\sum_{k'}\kappa_{k'(0)}G_{kk'};\\
\Delta_{k(jl)}=\sum_{k'\ne j,l}\kappa_{k'(jl)}G_{kk'};\\
\kappa_{k(0)}=u_{k(0)}v_{k(0)};\\
\rho_{k(0)}=\mid v_{k(0)}\mid^{2};\\
\kappa_{k(jl)}=u_{k(jl)}v_{k(jl)};\\
\rho_{k(jl)}=\mid v_{k(jl)}\mid^{2};\\
P_{0}=\mid c_{0}\mid^{2};\\
P_{jl}=\mid c_{jl}\mid^{2};\\
S_{0jl}=c_{0}c_{jl}^{*};\\
S_{mnjl}=c_{mn}c_{jl}^{*}.
\end{array}
\label{notatii}
\end{equation}
The symbol $\Delta_\gamma$ gives the gap parameter. (The label $\gamma$ denotes here
generically a specific configuration.) The variables that
depend on the time through the generalized coordinates are
the single particle densities $\rho_\gamma$, the pairing moment components $\kappa_\gamma$,
the probabilities to have a given seniority configuration $P_\gamma$,  and
the moment components between configurations $S_{\gamma\gamma'}$.
The relations (\ref{ec1})-(\ref{ec2}) are the well known
time dependent paring equations previously deduced in Refs. \cite{koonin,blocki}. 
These formulas are identical to the time dependent
Hartree-Fock-Bogoliubov equations \cite{avez,ebata}.
The symbol $h_{\gamma}$ denotes the Landau-Zener interaction, while $\bar{E}_{\gamma}$ 
and $T_{\gamma}$ are energy terms.
The significance of the quantities appearing in the equations can be
understood in the Appendix. The condition that $\sum_\gamma P_\gamma$=1
is implicitly ensured through Eqs. (\ref{ec3}) and (\ref{ec4})because $\dot{P}_0+\sum_{\gamma}\dot{P}_{\gamma}=0$. 

\subsection{The dissipation}

The energy of the nuclear system in the seniority zero state is
\begin{eqnarray}
\label{en0}
E_{0}=\langle \prod_k (u_{k(0)}(t)+v_{k(0)}(t)a_{k}^{+}a_{\bar{k}}^{+})\vert H
\vert \prod_k (u_{k(0)}(t)+v_{k(0)}(t)a_{k}^{+}a_{\bar{k}}^{+})\rangle\\
=2\sum_{k} \rho_{k(0)}\epsilon_{k}
- \sum_{k}\kappa_{(0)}\sum_{k'}\kappa_{k'(0)}^{*}G_{kk'} - \sum_{k}\rho_{k(0)}^{2}G_{kk};\nonumber
\end{eqnarray}
and in the seniority two state is
\begin{eqnarray}
\label{enij}
E_{jl}=\langle a_{j}^{+}a_{\bar{l}}^{+}\prod_{k\ne j,l}
(u_{k(jl)}(t)+v_{k(jl)}(t)a_{k}^{+}a_{\bar{k}}^{+}) \vert H \vert \\
\times
 a_{j}^{+}a_{\bar{l}}^{+}\prod_{k\ne j,l}
(u_{k(jl)}(t)+v_{k(jl)}(t)a_{k}^{+}a_{\bar{k}}^{+}) \rangle   \nonumber  \\
=2\sum_{k\ne j,l} \rho_{k(jl)}\epsilon_{k}
-\sum_{k\ne j,l}\kappa_{k(jl)}\sum_{k'\ne j,l}\kappa_{k'(jl)}^*G_{kk'} \nonumber \\
-\sum_{k\ne j,l}\rho_{k(jl)}^{2}G_{kk}+
\epsilon_{j}+\epsilon_{l}. \nonumber
\end{eqnarray}
The corresponding lower energy states $E^{0}_{0}$ and $E^{0}_{jl}$
 of the nuclear system in a given configuration
are obtained with the previous relations by replacing the densities $\rho_{\gamma}$
and the pairing moment components $\kappa_{\gamma}$ with the adiabatic values
obtained in the BCS approximation and using the same
single particle level scheme. Consequently, as defined in Ref. \cite{koonin}, along the fission path
the average dissipated energies $E^*_\gamma$ will be
\begin{equation}
\label{disip}
E^*_0=E_0-E^0_0;~~~
E^*_{jl}=E_{jl}-E^0_{jl},
\end{equation}
in the seniority zero and the seniority two configurations, respectively.
We subtracted from the total potential energy of the nuclear system,
its adiabatic value.
It was already shown in Ref. \cite{npa04} that the mean value of the dissipated energy becomes 
larger when the velocities of the
generalized coordinates increase.

The single particle levels belonging to the core of the initial parent
nucleus are rearranged in the two cores of the fragments. Knowing the
number of levels in each core, it is possible to redefine the pairing
active space of each fragment and the asymptotic values of the
lower energy states can be evaluated. 
After the scission, $E_0^0$ must be replaced by the sum
$E_{10}^0+E_{20}^0$, where $E_{10}^0$ and $E_{20}^0$ are the lower energies of
the two fission fragments. A similar rule is valid also for
seniority two configurations. In the same time, the pairing interaction matrix elements
$G_{kk'}$ between pairs pertaining to different nuclei are zero. 
For each channel,
asymptotically \cite{prc11}, the next limits hold
\begin{equation}
\label{disip2}
E^0_0\rightarrow E_{10}^0+E_{20}^0; ~~E_{0jl}^{0}\rightarrow E_{1jl}^0+E_{2jl}^0,
\end{equation}
where the indexes 1 and 2 refer to the two fragments.
If only one pair is broken along the fission path, 
the spin of the two nuclei delivered from one unique seniority two 
configuration must be the same.   This last assumption is due to the fact
that the Landau-Zener effect is produced between levels with the same good
quantum numbers. 

\subsection{The macroscopic-microscopic method and the single-particle energies}

In order to determine the fission barriers, the total energy of the nuclear system is computed in the
framework of the macroscopic-microscopic method
\cite{c4,c5}. As mentioned previously, the whole system is
characterized by some collective coordinates that determine
approximately the behavior of many other intrinsic variables.
The essential idea of this approach
is that a macroscopic model, as the liquid
drop one, describes quantitatively the smooth trends of the
potential energy with respect the particle number and deformation
whereas a microscopic approach as the shell model describes local
fluctuations. The combined macroscopic-microscopic method should
reproduce both smooth trends and local fluctuations.
The basic ingredient in such an analysis is the shape
parametrization that depends on several macroscopic degrees of
freedom. 
The macroscopic deformation energy is
calculated within the liquid drop model. A microscopic potential
must be constructed to be consistent with this nuclear shape
parametrization. A microscopic correction is then evaluated using
the Strutinsky procedure \cite{funny}.

\begin{figure}
\resizebox{0.45\textwidth}{!}{
  \includegraphics{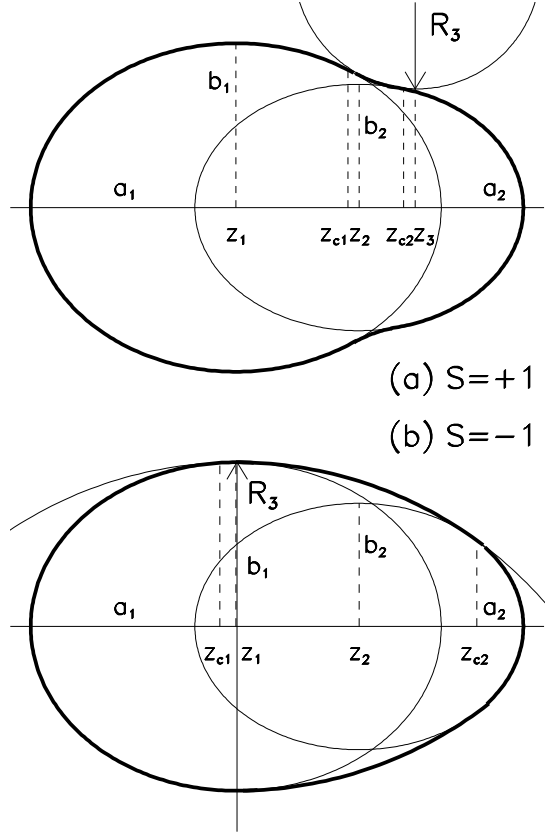}}
\caption
{Nuclear shape parametrization. The elongation is defined as $R=z_2-z_1$.
The curvature of the neck parameter is $C=S/R_3$, where $S=1$ for
necked shapes in the median surface and $S=-1$ otherwise. The eccentricities
of the fragments are $\epsilon_i=\sqrt{1-(b_i/a_i)^2}$ ($i=1,2$).
The mass asymmetry parameter can
be defined as $\eta=a_1b_1^2/(a_2b_2^2)$.
}
\label{figura2}
\end{figure}

The basic ingredient of the model is the nuclear shape parametrization.
In the following,
an axial symmetric nuclear shape surface during the deformation
process from one initial nucleus to the separated fragments is
obtained by smoothly joining two spheroids of semi-axis $a_i$ and $b_i$
($i$=1,2) with a neck surface generated by the rotation of a circle of radius $R_3$
around the axis of symmetry. By imposing the condition of volume
conservation we are left by five independent generalized
coordinates $\{q_n\}$ ($n$=1,5)
that can be associated to five degrees of freedom: the
elongation $R$ given by the distance between the centers of the
spheroids; the necking parameter $C=S/R_3$ related to the curvature of the
neck, the eccentricities $\epsilon_i$ associated with the deformations of the
nascent fragments and the mass asymmetry parameter $\eta=V_1/V_2$,
$V_i$ ($i=$1,2) denoting the volumes of the virtual
ellipsoids  characterized by the semi-axis $a_i$ and $b_i$.
The nuclear shape parametrization is displayed in Fig. \ref{figura2}.
The entire model can be considered valid as long as the 
generalized coordinates and their
variations in time make sense.

The many-body wave function and the single particle energies
are provided  by the  Woods-Saxon two-center shell model \cite{prc08}. The Woods-Saxon
potential, the Coulomb interaction and the spin orbit term must be diagonalized in an
eigenvector basis.
The asymmetric  two center shell oscillator provides an orthogonal eigenvector basis for only one Hermite space
\cite{prc96,npa2c}.
In this Hermite space the behavior of both fragments can be described.
When the elongation $R$ is zero, the eigenvector
basis becomes that of the anisotropic oscillator. When $R$ tends to infinity, a two oscillator
eigenvector system is obtained in the same Hermite space, 
centered in the two fragments. In the intermediate
situation, each eigenfunction has components in the two subspaces that belong to the fragments.
So, the two center shell model provides permanently the wave functions associated
to the lower energies of the single particle states pertaining to a major quantum number $N_{max}$.
Therefore, molecular states formed by two fragments at scission could be precisely described.
Another feature of the two center shell model is related to the
localization of the single particle wave function in one of the two
potential wells after the scission. As evidenced in Ref. \cite{prc11}, it
is possible to predict this localization for a given fragmentation before
that the scission is produced. This feature helps us to fix the
number of particles in each fragment by resolving Eqs. (\ref{ec1})-(\ref{ec6}).

This model was widely used by the Bucharest group
in the calculations
addressing the cluster \cite{m5,m8,m1,m4,mnou1} and alpha decay  \cite{alf},
the  fission \cite{m12,m10,mnou2},
or the heavy element synthesis \cite{m14}. For example, the model was able
to describe two tangent nuclei in a wide range of mass asymmetries. 
The half-lives for cluster decay were reproduced. A mechanism
for the formation of an $\alpha$-particle on the nuclear surface 
was supplied. Fission barriers that agree with the evaluated ones
were calculated.

If the different penetrabilities which characterize every channel
through the barrier are taken into account, it is expected that
the daughter ground state is strongly enhanced in the exit
channel. Indeed, in the cases of the excited channels, the
barrier must be increased with approximately the value of the
excitation energy of the unpaired nucleon and, therefore, the
penetrability is decreased exponentially. The amount of
which the barrier is modified can be estimated accounting the
specialization energy. Wheeler defined this specialization energy \cite{wheeler}
 as the excess of the energy of a nucleon with a given spin
over the energy for the same spin nucleon state of lowest
energy. So, the final excitation is given by
the sum between the dissipated energy and the single particle excitations that are
equivalent to the
 specialization energies.

The probability to obtain a given partition in the mass distribution is determined
by the penetrability of the fission barrier
\begin{equation}
P_b=\exp\left\{-{2\over\hbar}\int_{r_a}^{r_b}
\sqrt{B(r)[V(r)-U]}dr\right\}.
\end{equation}
The exponent is the classical action integral
taken along the fission path at a given energy $U$ that
connects the two turning points $r_a(U)$ and $r_b(U)$.
Only positive values of $[V(r)-U]$ are integrated. $B$ is the inertia
along the fission trajectory and $V$ is the deformation energy.

\section{Results}

\begin{figure}
\resizebox{0.85\textwidth}{!}{
  \includegraphics{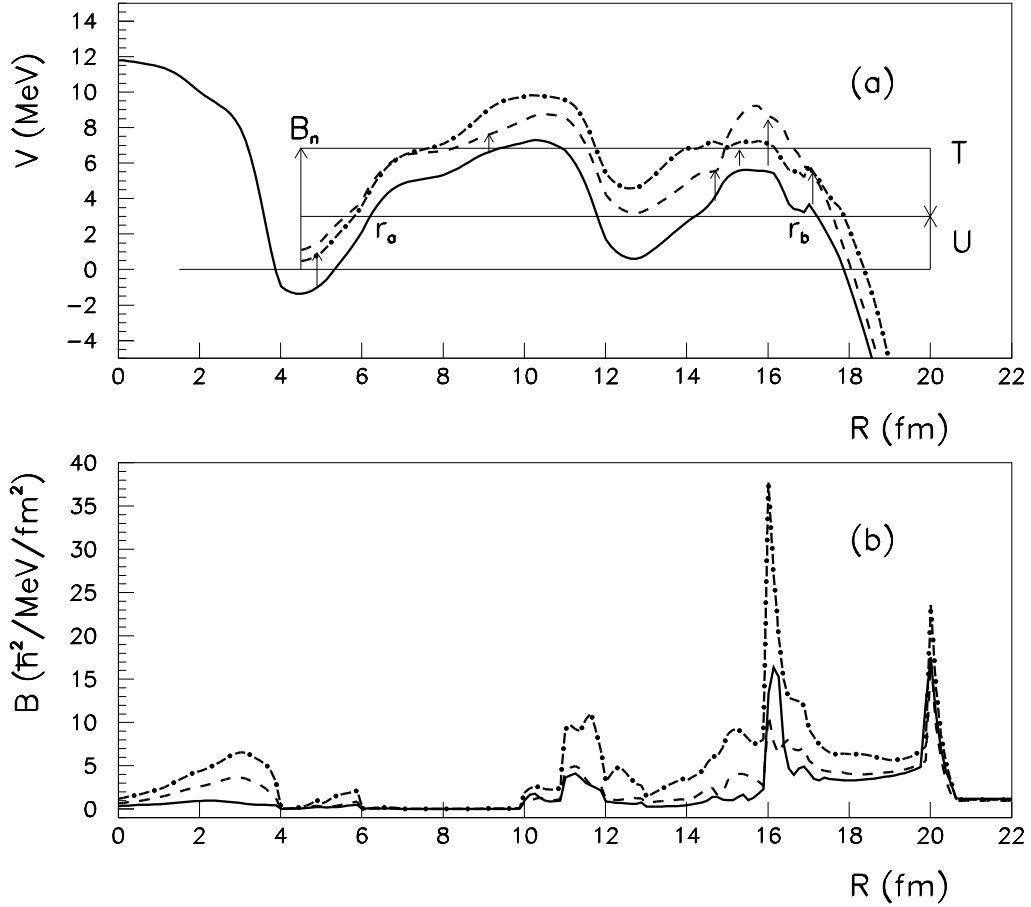}}
\caption
{(a) The the fission barrier is plotted with a full line.
The lower single particle excited seniority two state for neutrons is plotted
with a dot-dashed line while the first one 
for protons is plotted with a dashed line. The avoided level crossing regions are
marked with arrows. The turning points $r_a$ and $r_b$ are defined for
a given collective potential energy $U$.
(b) The inertia in the non adiabatic cranking approximation is plotted
with a full line, the inertia within the Gaussian overlap approximation 
is plotted with a dashed line and the inertia in the cranking approximation 
is plotted with a dot-dashed line.
}
\label{pot}
\end{figure}

\begin{figure}
\resizebox{0.85\textwidth}{!}{
  \includegraphics{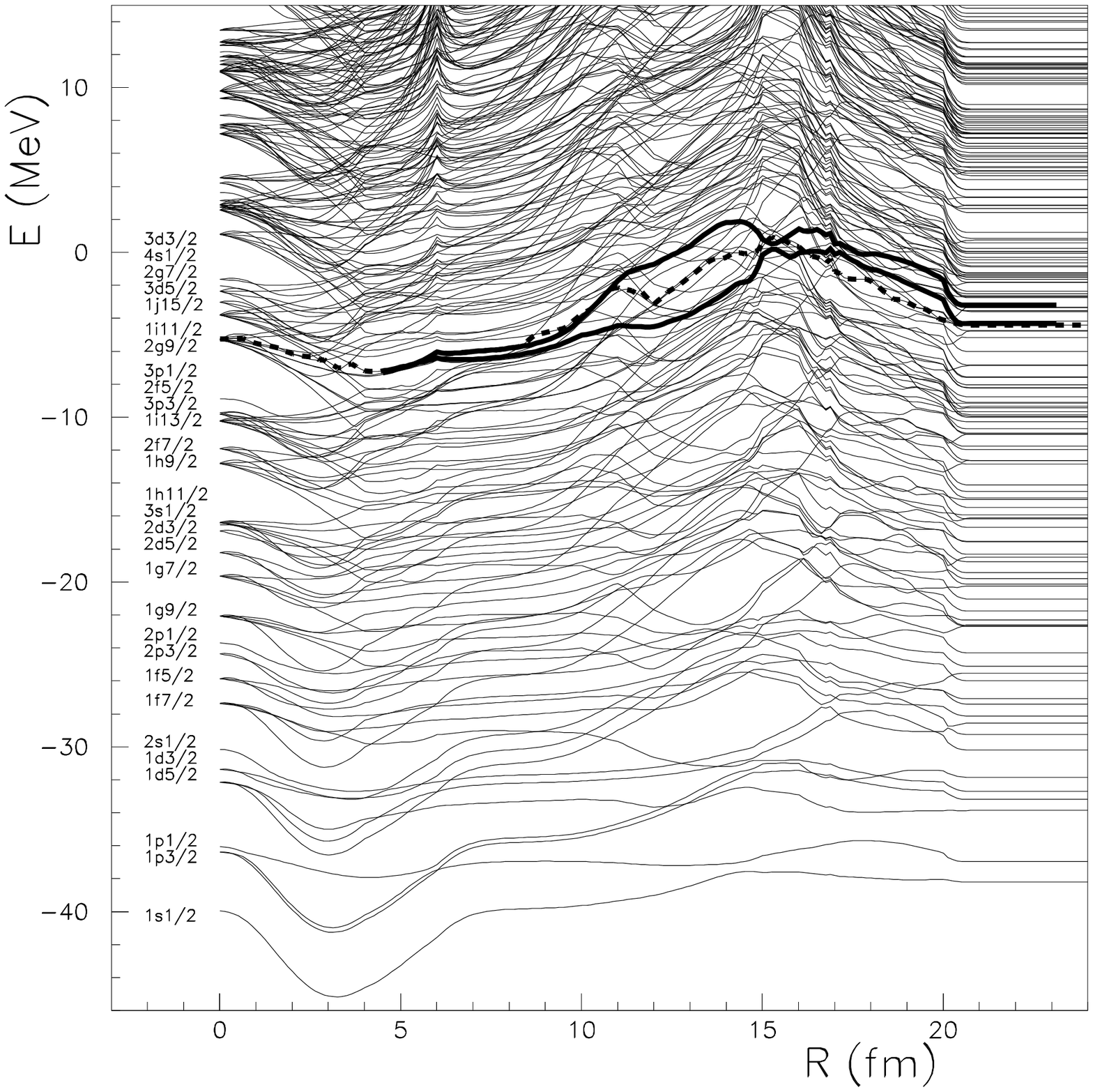}}
\caption
{Neutron single particle level scheme. 
The levels with spin projection 
$\Omega$=5/2 that give the lower energy configuration
for the unpaired fragments is plotted with a full line.
The Fermi energy  of the compound nucleus is displayed with a
thick dashed line.
Four avoided level crossing regions were identified for
$R\approx$ 4.9, 15.3, 17.1 and 19.6 fm.
}
\label{ttn}
\end{figure}

\begin{figure}
\resizebox{0.85\textwidth}{!}{
  \includegraphics{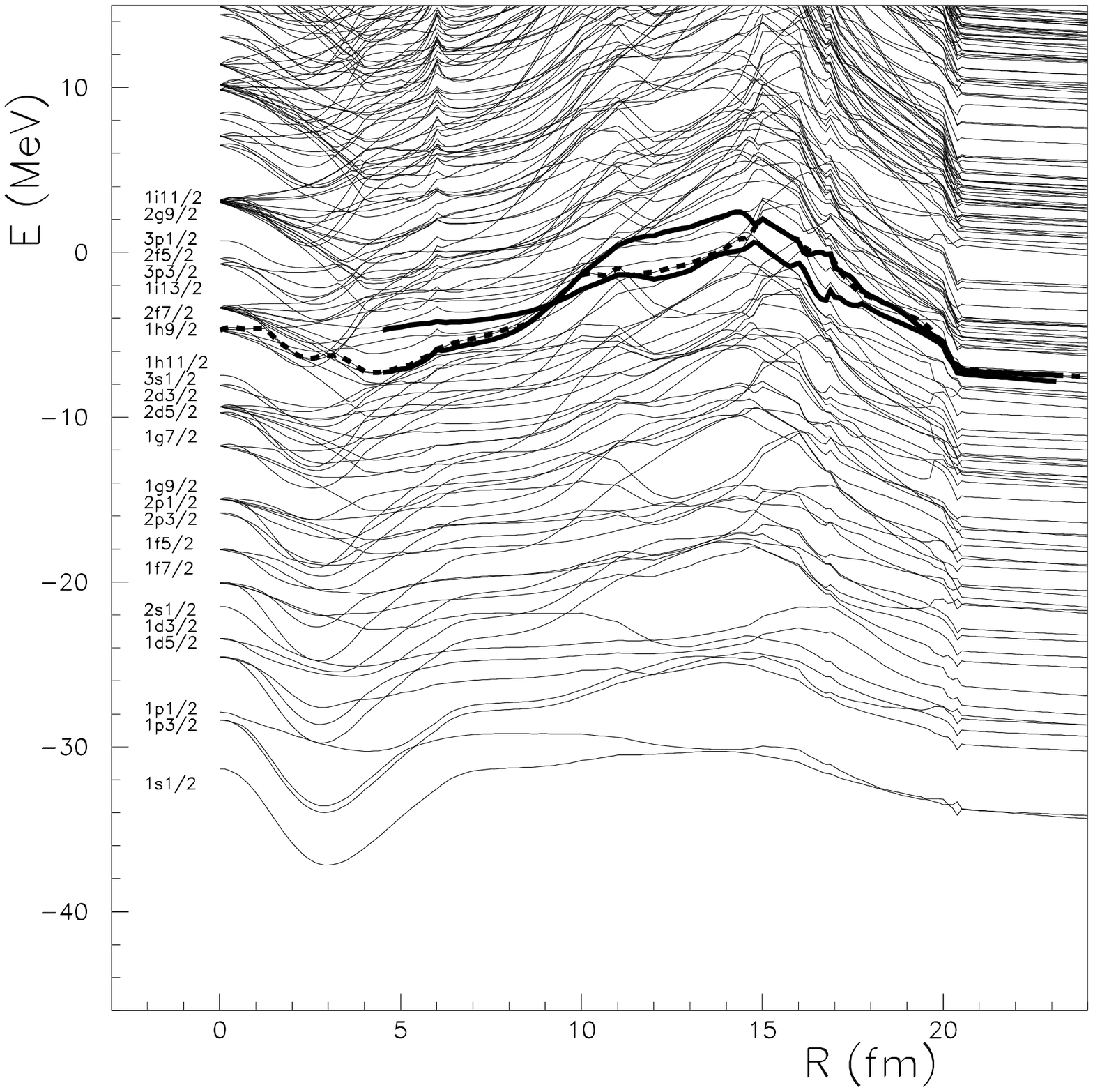}}
\caption
{Proton single particle level scheme. 
The levels with spin projection 
$\Omega$=3/2 that give the lower energy configuration
for the unpaired fragments is plotted with a full line.
The Fermi energy  of the compound nucleus is displayed with a
thick dashed line.
Four  avoided level crossing regions were identified for
$R\approx$ 9.1, 14.7, 16  and 20 fm.                    
}
\label{ttp}
\end{figure}

\begin{figure}
\resizebox{0.85\textwidth}{!}{
  \includegraphics{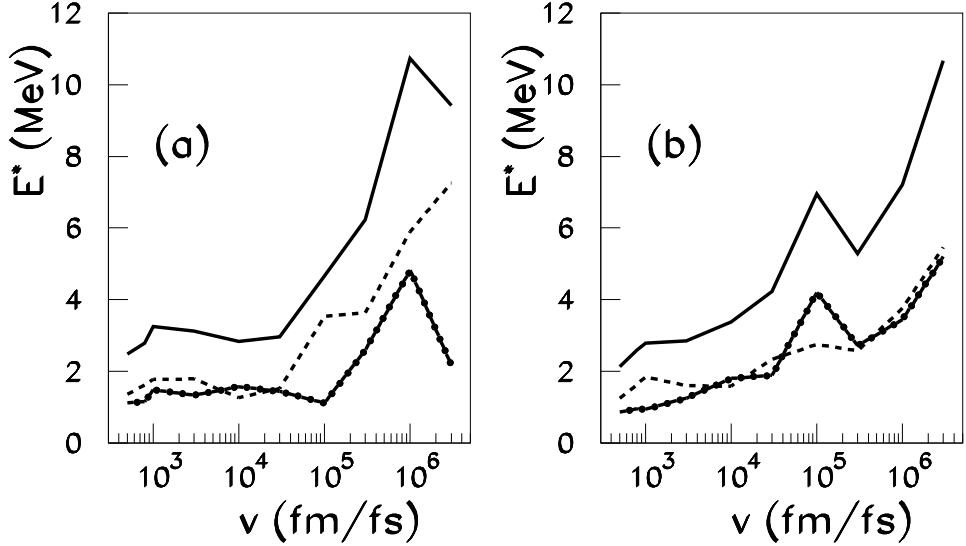}}
\caption
{ (a) Total dissipation energy after the scission 
$E^*$  as function of the internuclear
velocity $v$ for the seniority zero configuration. The neutron and proton
components of the dissipated energy are plotted with a dot-dashed and
a dashed line respectively.
(b) Total dissipation for the seniority two configuration
with lower single particle excitation. The neutron and proton components
are displayed with the same line types as in panel (a). 
}
\label{gre}
\end{figure}

\begin{figure}
\resizebox{0.85\textwidth}{!}{
  \includegraphics{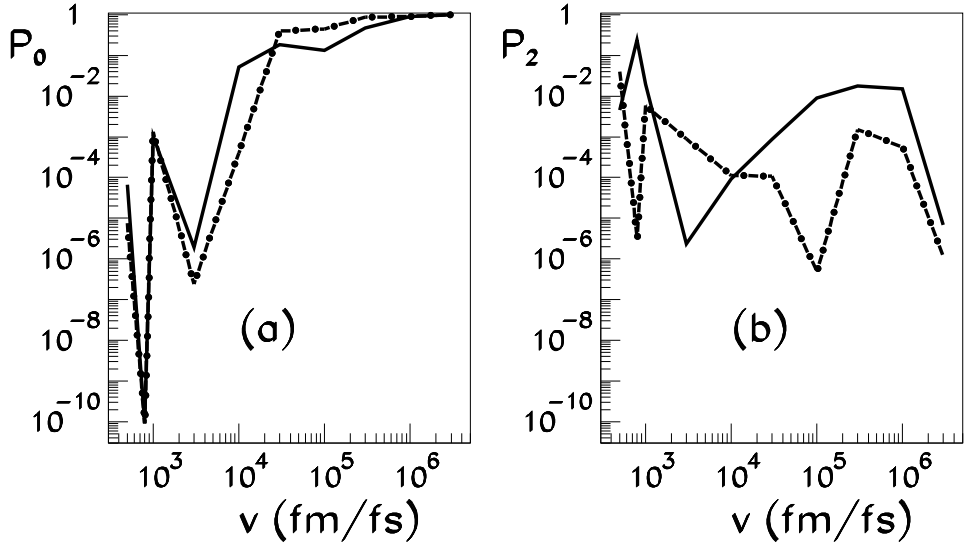}}
\caption
{ (a) Probability of realization of the adiabatic seniority zero
configuration at scission as function of the internuclear velocity for neutrons
(full line) and protons (dot-dashed line).
(b) Probability of the realization of the seniority two configuration 
at scission with the lower single particle excitation for neutrons (full line)
and protons (dot-dashed line).
}
\label{pro}
\end{figure}

\begin{figure}
\resizebox{0.85\textwidth}{!}{
  \includegraphics{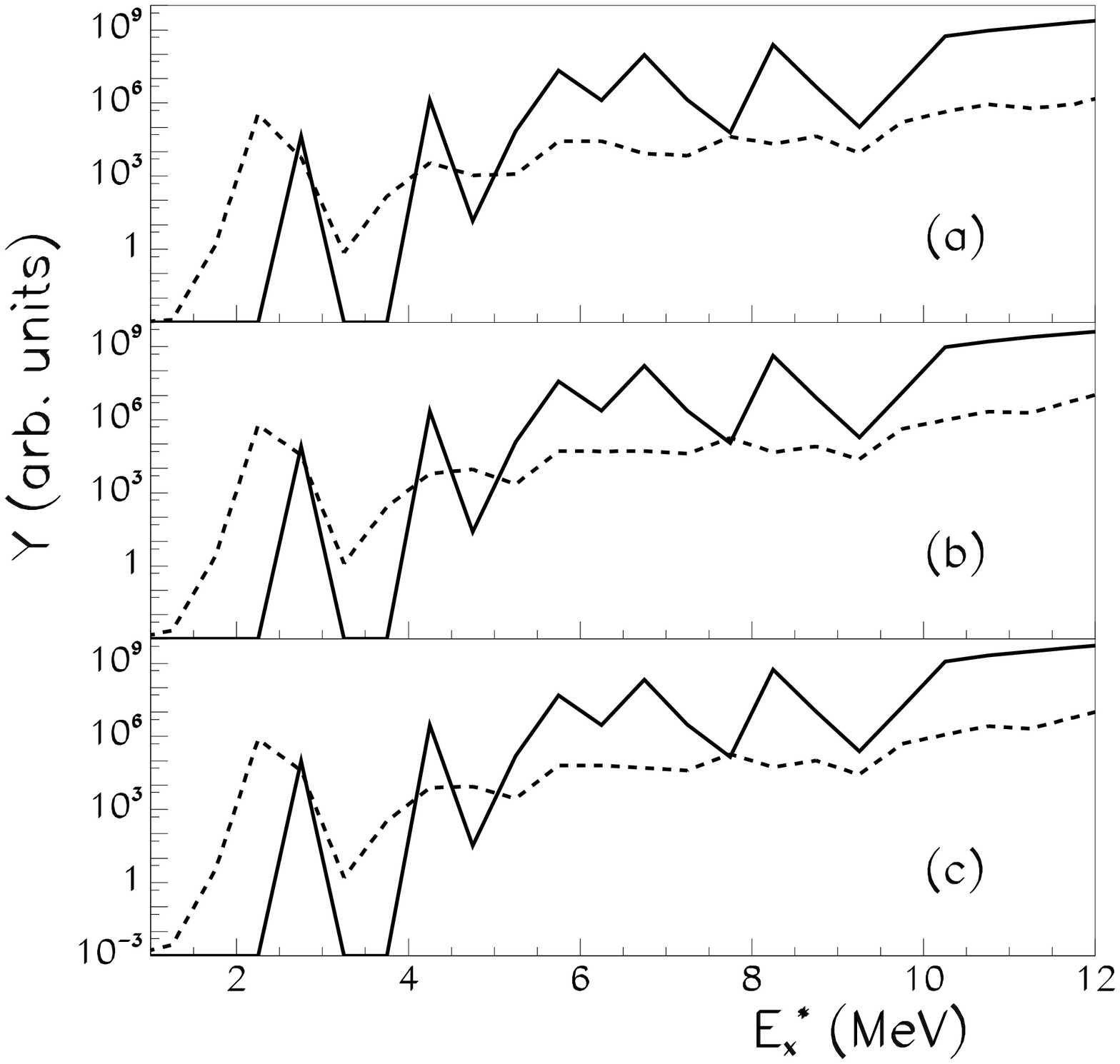}}
\caption
{ 
Full line: dependence of even even yield in arbitrary units $Y_0$ as function
of the final excitation $E_x^*$ of the fission fragments. Dashed line:
dependence of the odd-odd yields $Y_2$ as function of the excitation energy.
The panel (a) corresponds to the cranking model, the panel (b) is obtained
with the non-adiabatic cranking approach and the panel (c) with the Gaussian overlap approximation.
}
\label{grafinal}
\end{figure}

The fission yields as function of the total excitation energy for
the fragmentation $^{90}$Kr+$^{144}$Ba (even-even),
$^{90}$Rb+$^{144}$Cs (odd-odd) will be investigated in the
framework of our model. Experimental data are available for the 
reaction \cite{schwab} $^{233}$U(n$_{th}$,f)
and the model can be tested.

In order to determine the rearrangement of the single particle energies,
the first step is the calculation of the fission path from the ground
state up to scission. As described in Ref. \cite{m10},
the minimal action principle can be used to determine the best fission trajectory
in the configuration space spanned by our five generalized coordinates.
For this purpose, two ingredients are
required: the deformation energy $V$ and the tensor of the effective
mass $\{B_{ij}\}$. The deformation energy was obtained in the frame of
the microscopic-macroscopic method. 
The effective mass is computed within the
cranking approximation. By minimization, the theory can give
us the most probable fragmentation, that corresponds to a heavy fragment
with mass around 138. In our work, the
fission path for the 144/90 fragmentation is required. As evidenced in Ref. \cite{ran}, 
it is not yet understood how the compound nucleus is transformed
in a variety of different fragmentations. It is also
believed that the models for mass distributions have limited predictive
power. To overcome these difficulties, some simple assumptions
are made in our work. We rely on the results of
Refs. \cite{plb12,m12} where the mass distribution of the fragments was relatively
well reproduced by considering that the variation of the
mass asymmetry is linear from the saddle configuration of the
outer barrier up to the exit point. It is considered that under
the rapid descent from the top of the second barrier different
mass partitions are obtained. To obtain these different mass partitions,
the simplest way is to vary the mass asymmetry parameter
$\eta$ and the averaged deformations of fragments as in Ref. \cite{plb12}.
The fission barrier is plotted in Fig. \ref{pot} for the partition
$A_1/A_2=144/90$. At $R\approx$17 fm, the variation of the mass asymmetry
parameter is started, producing a small bumb in the barrier. In
the panel (b) of Fig. \ref{pot}, the effective mass is displayed.
The inertia along
the trajectory was computed within three approximations: the non-adiabatic cranking 
model \cite{jpg,mrrp}, the Gaussian overlap approximation  \cite{go1,go2},
and the cranking approach \cite{hill}.
In order to solve the equations
of motion, we need the single particle energies, the variations of the 
pairing matrix elements $G_{kl}$, the
interactions in the avoided crossing regions, 
and the velocity of
passage through these regions.

In Figs. \ref{ttn} and \ref{ttp}, the single particle energy levels obtained
within the two center Woods-Saxon model are displayed
along the fission trajectory for neutrons and protons, respectively. 
With a dashed thick line the Fermi levels
of the parent nucleus are plotted. The pair of levels with the same spin projection
that gives the lower single particle seniority two excitation (or
specialization energy) are also
displayed with thick full lines. It can be observed that the levels
that pertain to the lower excitation are close to the Fermi energy.
In the case of protons the spin projection $\Omega$ of these levels is 3/2 while
in the case of neutrons it is $\Omega$=5/2. These values of $\Omega$ give
also the final spin of the partners. The barriers for the seniority two
configurations were displayed in Fig. \ref{pot}
with dot-dashed and dashed lines. The difference
between the seniority two and seniority zero
energies, that is the single particle excitations, 
are obtained by subtracting their  values 
 computed with Rels. (\ref{enij}) and (\ref{en0}) and
by using the adiabatic BCS amplitudes.
Our pairing active space is constructed with 58 single particle
levels around the Fermi energy.

An important question is the identification of the avoided
crossing regions that arise from the strong energy fluctuations of the
single particle levels observed
along the fission path. For this purpose, we selected 
all the pairs of adjacent levels with the same
spin projection in the active pairing space. From an analyze of the
rearrangement of these levels in a manner similar to that of Ref. \cite{mirt}, 
the avoided crossing regions  were identified. The positions of some
 avoided level crossing
regions are marked with arrows in Fig. \ref{pot}. In these regions, the
excitations are possible between the mentioned seniority two configurations and
the seniority zero one.
Using the interpolation method described in Ref. \cite{greinerscheid},
the magnitude of the interactions was calculated. Finally, 
we selected 32 seniority two configurations for protons and 31 configurations for neutrons
that are coupled to the seniority zero configuration through 
avoided levels crossings regions. In the frame of the adiabatic BCS model \cite{hill},
a mean value of the pairing interaction $G$ can be associated to a given active
pairing space, by using a renormalization procedure.
So, the values of $G$ for all seniority states were computed for the parent and for the 
two fragments.
In the case of seniority two states, the blocked levels are eliminated from the single
particle diagram.
A linear interpolation between the values of $G$ from those of the
parent to those of the fragments is realized in order to solve the equations of motion,
as it will be specified in the following. 
So, all the most important quantities required to
solve the equations of motion are provided: the single particle energies $\epsilon_k$,
the perturbations $h_{ij}$ and the pairing interactions $G$.
The dependence in time is introduced by means of the variations
of the collective coordinates. In this respect,
several values of the internuclear velocity $v=\dot{R}$ are tested in order to
solve the equations: 5$\times 10^2$, 8$\times 10^2$, $10^3$, 3$\times 10^3$,
$10^4$, 3$\times 10^4$, $10^5$, 3$\times 10^5$, $10^6$, and 3$\times 10^6$  fm/fs.

The Eqs. (\ref{ec1})-(\ref{ec2}) don't have a dependence on
$P_{\gamma}$ and $S_{\gamma\gamma'}$ and can be solved separately for different
values of the internuclear velocities. The initial conditions are given by 
the adiabatic BCS values for the ground state deformation of the parent 
(located at $R\approx$ 4.6 fm) of all seniority states involved. 
The equations are solved in a way similar to that presented in Ref. \cite{prc11}.
That is, at the beginning of the reaction the system evolves without constraints. 
 At the internuclear distance $R$=17 fm, close
to the top of the outer barrier, in Eqs. (\ref{ec22}) and (\ref{ec2})
the value of $\lambda$ is set to non-zero
values. 
In the same time, two linear variations of the mean value of the pairing
interactions $G$  are started in order
to reach the final values $G_1$ and $G_2$ that characterize the two fragments at scission.
When the equalities between the sums of single particle occupation probabilities
and the numbers of nucleons are obtained,
$\lambda$ and the pairing interaction between the two fragments
$G_{12}$ are set to zero. For $G_{12}=0$, from conservation conditions as
explained in Ref. \cite{prc11}, the sum of occupation probabilities that run
on the single particle states  of the two fragments are preserved.
The solutions $\rho_{k(\gamma)}$ and $\kappa_{k(\gamma)}$ 
are obtained up to an internuclear distance of $R$=22 fm. 
The calculated values of $\bar{E}_{\gamma}$ and
$T_{k\gamma}$ as function of $R$ are recorded and will 
be used later to solve the Eqs. (\ref{ec3})-(\ref{ec6}).
For the two isospins and for all velocities and  seniority 
configurations involved, the systems are solved within the Runge-Kutta algorithm.
The dissipated energies $E^*$ after the scission as function of the internuclear velocity $v$, 
computed according to Eqs. (\ref{disip}) and (\ref{disip2}),
are plotted in Fig. \ref{gre} for the seniority zero and the lower single particle excitation 
seniority
two states. The general trend exhibited
in both cases is an increase of the final dissipated energy
when the internuclear velocity becomes larger. The same behavior is typical for all seniority 
configurations involved. So the general rule that
assesses that the dissipation is proportional to the velocity is retrieved.
Another way to introduce the dissipation in quantum systems is to consider forces proportional
to the velocity in analogy with the friction forces in classical mechanics \cite{scut},
leading to a generalized Schr\"odinger non-linear equation for an open system.

The initial condition for the equations (\ref{ec3})-(\ref{ec6}) that describe the configuration
mixing is $P_0$=1, all the other values being zero.
From physical considerations, we imposed in the numerical code
the conditions
that $\dot{P}_{ij}\le$0 if $P_{ij}=1$ or if $P_0=0$, and
that $\dot{P}_{ij}\ge$0 if $P_{ij}=0$ or if $P_0=1$. That
ensures that the probability to have any configuration must be in
the interval $[0,1]$.  The final
probabilities of realization of two configurations for 
 neutrons and protons as function of the internuclear velocities are 
displayed in Fig. \ref{pro}.
This figure evidences the fact that the probability of realization
of adiabatic seniority zero states are close to zero for small velocities, exactly
in the energy region in which the dissipation is smaller.
So, the main features concerning the relation
between the final excitation energy and the probability
of realization of a given channel presented in Ref. \cite{plb09} are retrieved
and are valid even if the method for dynamical projection on final atomic and
mass numbers is used.

The results concerning the final excitation and the probabilities of realization
presented above
are coupled through the collective velocity parameter. In order
to compare the theoretical findings with the experimental data, this
velocity must be eliminated. Therefore,
a simple model related to the experimental arrangement
is conceived in order to relate the dissipation
and the yields. By bombarding the $^{233}$U with thermal neutrons,
using mass evaluations \cite{audi} we found that an excitation
of at least $B_n$=6.85 MeV is accumulated in the compound $^{234}$U nucleus.
As represented in Fig. \ref{pot},
this energy is shared between a potential part $U$ and a kinetic one
$T=2B_0v^2/2$, where $B_0$ is the inertia in the ground state
configuration. A constant population of all values of the
kinetic energy is assumed. The penetrability $P_b(v,\gamma_n,\gamma_p)$
of each channel depends on the turning
points of the barrier, that is, on $v$ through the relation $B_n=U+T$. 
That leads to a larger penetrability when the velocity decreases.
The penetrability depends also on the excitation channel $\{\gamma\}$ that can
be constructed with the specialization energies of the 
configurations $\{0\}$ or $\{ij\}$
of the two isospins.
The final excitation energy of the two fragments is 
$E^*_x(v,\gamma_n,\gamma_p)=E^*(v,\gamma_n,\gamma_p)+E_{sp}(\gamma_n,\gamma_p)$,
that is, it corresponds to a sum between the dissipated energy
$E^*$ and the single particle excitations $E_{sp}$ of the two fragments.
The dependence of the yield as function of the final excitation $E_x^{*}$ in the
interval [$E_{x1}^*,E_{x2}^*$] exhibits the following proportionality:
\begin{eqnarray}
\label{y0}
Y_0(E_x^*)\propto {1\over E_{x2}^*-E_{x1}^*}\int_{0}^{v_{max}}
[P_{0(n)}(v)P_{0(p)}(v)P_b(v,0,0)\\
+P_{0(n)}(v)\sum_{\gamma_p}P_{\gamma_p(p)}(v)P_b(v,0,\gamma_p)\nonumber\\
+P_{0(p)}(v)\sum_{\gamma_n}P_{\gamma_n(n)}(v)P_b(v,\gamma_n,0)\nonumber \\
+\sum_{\gamma_n,\gamma_p}P_{\gamma_n(n)}(v)P_{\gamma_p(p)}(v)P_b(v,\gamma_n,\gamma_p)]
\nonumber\\
\times w(v)\theta(E_{x1}^*-E_x^*(v,\gamma_n,\gamma_p))\theta(E_x^*(v,\gamma_n,\gamma_p)-E_{x2}^*))dv,\nonumber
\end{eqnarray}
for the even-even channel. Here $P_b$ are the penetrabilities that depend
on a specific channel and $P_0$ are the probabilities of realization given
by the time dependent equations for
neutrons (index $(n)$) and protons (index $(p)$). For all velocities
that give an excitation in the interval [$E_{x1}^*,E_{x2}^*$], the 
previous formula reflects the fact that the yields are proportional to
these penetrabilities and probabilities.
Of course, at scission it is possible to obtain even-even
partitions even if a Cooper pair is broken. In this case, one of the
fission products picks this broken pair and will carry a very large
excitation energy.
Therefore, the sums that run over the channels $\gamma$ in the
expression (\ref{y0}) take into consideration the fact that some
configurations are formed with a broken pair in only one partner.
The probabilities of realization of a given seniority configuration $P_{\gamma_n(p)}(v)$
 depend on the internuclear velocity.
The penetrability of the fission barrier $P_b(v,\gamma_n,\gamma_p)$  at the velocity $v$
in the channel $\{\gamma_n,\gamma_p\}$ depends on the variation of the
probabilities of realization along the fission path. The factor
$w(v)=B_0v$ is a weighting that reflects the dependence of the kinetic energy
of the velocity, because $dT=w(v)dv$, and $\theta$ is the step Heaviside distribution
used to select only events in the interval $[E_{x1}^*,E_{x2}^*]$.
$B_0$ is considered to be the inertia in the ground state of the parent nucleus. 
The maximal value of the the velocity is obtained from the boundary  $T_{max}={1\over 2}B_0v_{max}^2=B_n$.
The penetrability $P_b(v,\gamma_n,\gamma_p)$ is calculated 
by considering the turning points at the energy
$U=B_n-{1\over 2}B_0v^2$. 
The calculations were made within three approximations
for the inertia: the cranking model \cite{hill}, the Gaussian overlap approximation \cite{go1,go2}
and the non-adiabatic cranking approach \cite{jpg,mrrp}.
The yields for odd-odd partitions are proportional
with the expression:
\begin{eqnarray}
Y_2(E_x^*)\propto {1\over E_{x2}^*-E_{x1}^*}\int_{0}^{v_{max}}\sum_{\gamma_n}\sum_{\gamma_p}
P_{\gamma_n(n)}(v,)P_{\gamma_p(p)}(v,)P_b(v,\gamma_n,\gamma_p)w(v)\nonumber\\
\times\theta(E_{x1}^*-E_x^*(v,\gamma_n,\gamma_p))\theta(E_x^*(v,\gamma_n,\gamma_p)-E_{x2}^*)dv.
\end{eqnarray}
In the previous relation, the sums run over all the seniority two configuration taken into consideration for neutrons
(index $n$) and protons (index $p$) that give unpaired nucleons in the two fragments.

The results obtained for the folded  distributions $Y_0(E_x^*)$ and $Y_2(E_x^*)$ are displayed in Fig. \ref{grafinal}.
The averaging interval is $E_{x2}^*-E_{x1}^*$=0.5 MeV.
The observed experimental trends exhibited in Fig. 4 of Ref. \cite{schwab} for cold fission yields
were reproduced. The trends are same for the three approaches used for the inertia.
At low excitation energy the odd-odd yields surpass the even-even ones. The even-even yields
become larger for excitation energies larger that 3-4 MeV, in accordance with the experimental
findings.

\section{Discussion and conclusions}

In this work, a microscopic model is proposed for the explanation of the
odd-even effect in cold fission. This explanation is based on  a mixing configuration
mechanism that is produced during the fission process. This configuration
mixing mechanism is obtained dynamically by solving a 
the generalized system of
time dependent pairing equations, that include a pair-breaking effect.
A first rule can be extracted from this model. The even-even fission
products cannot be obtained at zero excitation energies because
of the existence of dynamical excitations produced in the avoided level crossing regions
when the nuclear system deforms slowly.

The magnitudes of the interactions and the location of the
the avoided crossing regions are fixed along the fission path and
are independent of the velocity of passage through these regions.
If this velocity is large, the perturbation will act onto the Cooper
pair a small fraction of time. So, the chance to break a pair
will be small. If the velocity is low, the pairs will traverse
the regions in larger time durations. So, the probability to break
a pair increases. On another hand, high velocities lead to
large dissipation energies.

Another characteristic that was not exploited in this work
can be featured from the model.
The lower excitation energies of a combination of two odd-odd partners
can be obtained only if their spin are the same for neutrons
and protons. If the
spins of the partners are not the same, the model predicts that at least two
pair ruptures are produced for neutrons or protons and
additional single particle excitations must be taken into consideration.

The possibility to jump from one level to another in a large scale
amplitude motion was 
predicted by Hill and Wheeler in Ref. \cite{hill}. Dissipation in 
terms of Landau-Zener crossings during fission was first 
 proposed in Ref. \cite{sch} where excitations were considered 
 only for time-reversed pairs, neglecting the possible existence 
 of unpaired nucleons. As evidenced in Ref. \cite{hasse},
many studies were performed in order to exploit this
mechanism in different type of processes. It was also
shown in Ref. \cite{blocki}
that the Landau-Zener mechanism is cached in the time
dependent pairing equations (\ref{ec1}) and (\ref{ec2}).
Pairs undergo Landau-Zener transitions on virtual levels
with coupling strengths given by the value of the 
magnitude of the gap parameter.
Anyhow,
it is the first time that the dynamical pair breaking effect
was used to explain the odd-even effect in fission.
It must be mentioned that a time-dependent microscopic approach
to the scission process was described in Ref. \cite{rizea}.
They observed that for scission time larger than 5$\times10^{-21}$ s,
the single particle excitations are negligible. This time that characterizes
the neck rupture corresponds to a velocity of $10^6$ fm/fs in our
calculations.

The density of single particle levels increases in the region of the
second barrier. Therefore, from the outer saddle to the scission, many
 avoided crossing regions are produced and the chance to break a pair
is enhanced. Up to the second saddle, the number of avoided crossing regions
is small an the system evolves merely in the seniority zero configuration.
That gives a large penetrability for all channels, the mixing of
configurations being produced especially in the outer barrier region.

In conclusion, by solving the dynamical microscopic equations of motion
for an fissioning even-even system it is found that the probability
to obtain an odd-odd partition overcomes the probability of an even-even one
at excitation energies smaller than 4 MeV, for the same division in
mass numbers. The theoretical results are in accordance with the experimental
behavior of the odd-even distributions at high kinetic energies. It
is the first time that this behavior was explained within a quantum
mechanical approach.

{\bf{Acknowledgements}}

Work supported by CNCS-UEFISCDI, project number
PN-II-ID-PCE-2011-3-0068.

\appendix
\section{The equations for the configuration mixing}
\label{appendix1}

The next identities are used to develop the functional (\ref{var})
\begin{equation}
\begin{array}{c}
\left \langle c_{0}\prod_{k}(u_{k(0)}+v_{k(0)}a_{k}^{+}a_{\bar k}^{+})\left\vert
H(t)-\lambda\mid N_2\hat{N}_1-N_1\hat{N}_2\mid \right\vert\right.\\
\left.\times  c_{0}\prod_{k}(u_{k(0)}+v_{k(0)}a_{k}^{+}a_{\bar k}^{+})\right\rangle
\\=
\mid c_{0}\mid^{2}\left(2\sum_{k}\mid v_{k(0)}\mid^{2}(\epsilon_{k}-sN_{i_k}\lambda)\right. \\
\left. +\sum_{k}u_{k(0)}v_{k(0)}\sum_{k'}u^*_{k'(0)}v^*_{k'(0)}G_{kk'}
-\sum_{k}\mid v_{k(0)}\mid^{4}G_{kk}\right);
\end{array}
\end{equation}

\begin{equation}
\begin{array}{c}
\left\langle\sum_{j,l\ne j}c_{jl}a_{j}^{+}a_{\bar l}^{+}\prod_{k\ne j,l}
(u_{k(jl)}+v_{k(jl)}a_{k}^{+}a_{\bar k}^{+})\left\vert
H(t)\right.\right.
\\
\left.\left. -\lambda\mid N_2\hat{N}_1-N_1\hat{N}_2\mid
\right\vert \sum_{j,l\ne j}c_{jl}a_{j}^{+}a_{\bar l}^{+}\prod_{k\ne j,l}
(u_{k(jl)}+v_{k(jl)}a_{k}^{+}a_{\bar k}^{+})\right\rangle
\\=
\sum_{j,l\ne j}\mid c_{jl}\mid^{2}\left(2\sum_{k\ne j,l}
\mid v_{k(jl)}\mid^{2}(\epsilon_{k}-sN_{i_k}\lambda)\right.\\
+(\epsilon_{j}-sN_{i_j}\lambda)
+(\epsilon_{l}-sN_{i_l}\lambda)\\
\left. +\sum_{k\ne j,l}u_{k(jl)}v_{k(jl)}\sum_{k'\ne j,l}u^*_{k'(jl)}v^*_{k'(jl)}G_{kk'}-
\sum_{k\ne jl}\mid v_{k(jl)}\mid^{4}G_{kk}\right)
\end{array}
\end{equation}
where we introduced the sign
$s={\rm sign} (N_2\sum_{k_1}\mid v_{k_1(\gamma)}\mid^2-N_1\sum_{k_2}\mid v_{k_2(\gamma)}\mid^2)$
in order to have a positive value of the matrix element of the condition (\ref{coop}).
In this last relation $\{\gamma\}=\{0\}$ or $\{ ij\}$ denotes a configuration. $N_{i_k}=N_2$ 
or $N_{i_k}=-N_1$ if the state $k$ will belong to the fragment 1 or 2 after the scission,
respectively.

For the time derivatives, the next relations are used. 

\begin{equation}
\begin{array}{c}
\left\langle c_{0}\prod_{k}(u_{k(0)}+v_{k(0)}a_{k}^{+}a_{\bar k}^{+})\left\vert {\partial 
\over \partial t}
\right\vert c_{0}\prod_{k}(u_{k(0)}+v_{k(0)}a_{k}^{+}a_{\bar k}^{+})\right\rangle\\
=c_{0}^{*}\dot{c}_{0}+\mid c_{0}\mid^{2}\sum_{k}(u_{k(0)}\dot{u}_{k(0)}+
v_{k(0)}^{*}\dot{v}_{k(0)});
\end{array}
\end{equation}

and

\begin{equation}
\begin{array}{c}
\left\langle\sum_{j,l\ne j}c_{jl}a_{j}^{+}a_{\bar l}^{+}\prod_{k\ne j,l}
(u_{k(jl)}+v_{k(jl)}a_{k}^{+}a_{\bar k}^{+})\left\vert
{\partial
\over \partial t}\right\vert\right.\\
\left.\times
 \sum_{j,l\ne j}c_{jl}a_{j}^{+}a_{\bar l}^{+}\prod_{k\ne j,l}
(u_{k(jl)}+v_{k(jl)}a_{k}^{+}a_{\bar k}^{+})\right\rangle
\\=\sum_{j,l\ne j}\left[c_{jl}^{*}\dot{c}_{jl}+
\mid c_{jl}\mid^{2}\sum_{k\ne j,l}(u_{k(jl)}\dot{u}_{k(jl)}+
v_{k(jl)}^{*}\dot{v}_{k(jl)})\right].
\end{array}
\end{equation}
The matrix elements of the time derivatives of the wave functions
are neglected because they are considered to be responsible only
for the inertia parameter. These matrix elements were investigated in Refs.
\cite{jpg,mrrp}, where a relation between the inertia and the dissipation was
revealed.
By using the following properties of the creation and annihilation operators:
\begin{equation}
\label{proper}
\begin{array}{c}
\alpha_{k}a_{k}^{+}=u_{k}+v_{k}a_{k}^{+}a_{\bar k}^{+};~~
\alpha_{k}^{+}a_{\bar k}^{+}=u_{k}a_{k}^{+}a_{\bar k}^{+}-v_{k}^{*},\\
\alpha_{\bar k}a_{\bar k}^{+}=u_{k}+v_{k}a_{k}^{+}a_{\bar k}^{+};~~
\alpha_{\bar k}^{+}a_{k}^{+}=-u_{k}a_{k}^{+}a_{\bar k}^{+}+v_{k}^{*},
\end{array}
\end{equation}
the next equalities are deduced:
\begin{equation}
\begin{array}{c}
\left\langle c_{0}\prod_{k}(u_{k(0)}+v_{k(0)}a_{k}^{+}a_{\bar k}^{+})\left\vert
\sum_{i,j\ne i}\alpha_{i(0)}\alpha_{\bar j(0)} \right.\right.\\
\left.\left. \times \prod_{k\ne i,j}\alpha_{k(0)}a_{k}^{+}a_{k}\alpha^{+}_{k(ij)}
\right\vert \sum_{j,l\ne j}c_{jl}a_{j}^{+}a_{\bar l}^{+}\prod_{k\ne j,l}
(u_{k(jl)}+v_{k(jl)}a_{k}^{+}a_{\bar k}^{+})\right\rangle\\
=\sum_{i,j\ne i}c_{0}^{*}c_{ij};
\end{array}
\end{equation}
 
and

\begin{equation}
\begin{array}{c}
\left\langle \sum_{j,l\ne j}c_{jl}a_{j}^{+}a_{\bar l}^{+}\prod_{k\ne j,l}
(u_{k(jl)}+v_{k(jl)}a_{k}^{+}a_{\bar k}^{+})\mid
\sum_{i,j\ne i}\alpha_{i(0)}^{+}\alpha_{\bar j(0)}^{+} \right.\\
\left. \times \prod_{k\ne i,j}\alpha_{k(ij)}a_{k}^{+}a_{k}\alpha^{+}_{k(0)}
\mid c_{0}\prod_{k}(u_{k(0)}+v_{k(0)}a_{k}^{+}a_{\bar k}^{+})\right\rangle \\
=\sum_{i,j\ne i}c_{0}c_{ij}^{*}.
\end{array}
\end{equation}

Using  the previous identities,  the energy functional (\ref{var}) reads, eventually:
\begin{equation}
\begin{array}{c}
\left\langle \varphi \mid H-i\hbar{\partial\over \partial t}+H'-\lambda 
(N_2\hat{N}_1-N_1\hat{N}_2)\mid \varphi \right\rangle \\
=\mid c_{0}\mid^{2}\left(2\sum_{k} \mid v_{k(0)}\mid^{2}(\epsilon_{k}-sN_{i_k}\lambda)\right.\\
\left. -
\sum_{k}u_{k(0)}v_{k(0)}\sum_{k'}u^*_{k'(0)}v^*_{k'(0)}G_{kk'}
-\sum_{k}\mid v_{k(0)}\mid^{4}G_{kk}\right)\\+
\sum_{j,l\ne j}\mid c_{jl}\mid ^{2} \left\{2\sum_{k\ne j,l} \mid v_{k(jl)}\mid^{2}
(\epsilon_{k}-sN_{i_k}\lambda)\right.\\
+(\epsilon_{j}-N_{i_j}\lambda)
+(\epsilon_{l}-sN_{i_l}\lambda)\\
\left. - \sum_{k\ne j,l}u_{k(jl)}v_{k(jl)}\sum_{k'\ne j,l}u^*_{k'(jl)}v^*_{k'(jl)}G_{kk'}
- \sum_{k\ne j,l}\mid v_{k(jl)}\mid^{4}G_{kk}\right\}\\
-i\hbar\left\{c_{0}^{*}\dot{c}_{0}+\mid c_{0}\mid^{2}\sum_{k}{1\over 2}(v_{k(0)}^{*}\dot{v}_{k(0)}
-\dot{v}_{k(0)}^{*}v_{k(0)})\right. \\
\left. +\sum_{j,l\ne j}\left[c_{jl}^{*}\dot{c}_{jl}+\mid c_{jl}\mid^{2} \sum_{k\ne j,l}
{1\over 2}(v_{k(jl)}^{*}\dot{v}_{k(jl)}
-\dot{v}_{k(jl)}^{*}v_{k(jl)})\right]\right\}\\
+\sum_{l,j\ne l}^{n}h_{jl}(c_{0}^{*}c_{jl}+c_{0}c_{jl}^{*})\\
=\mid c_{0}\mid^{2}\bar{E}_{0}+\sum_{j,l\ne j}\mid c_{jl}\mid ^{2}\bar{E}_{jl}\\
-\left\{i\hbar c_{0}^{*}\dot{c}_{0}+\mid c_{0}\mid^{2}\sum_{k}T_{k(0)}
+\sum_{j,l\ne j}\left[i\hbar c_{jl}^{*}\dot{c}_{jl}+\mid c_{jl}\mid^{2} \sum_{k\ne j,l}
T_{k(jl)}\right]\right\}\\
+\sum_{l,j\ne l}^{n}h_{jl}(c_{0}^{*}c_{jl}+c_{0}c_{jl}^{*}),
\end{array}
\label{expr0}
\end{equation}
where $\bar{E}_0$ and $\bar{E}_{jl}$ are terms that include the energies $E_0$
and $E_{jl}$ (given by formulas (\ref{en0}) and (\ref{enij}))
of the seniority zero and the seniority two configurations, respectively,
\begin{equation}
\begin{array}{c}
\bar{E}_{0}=2\sum_{k} \mid v_{k(0)}\mid^2 (\epsilon_{k}-sN_{i_k}\lambda) \\
- \sum_{k}u_{k(0)}v_{k(0)}\sum_{k'}u^*_{k'(0)}v^*_{k'(0)}G_{kk'} - \sum_{k}\mid v_{k(0)}\mid^4G_{kk}\\
=E_0-2\sum_k \rho_{k(0)}sN_{i_k}\lambda;
\end{array}
\end{equation}

\begin{equation}
\begin{array}{c}
\bar{E}_{jl}=2\sum_{k\ne j,l} \mid v_{k(jl)}\mid^2(\epsilon_{k}-sN_{i_k}\lambda)\\
-\sum_{k\ne j,l}u_{k(jl)}v_{k(jl)}\sum_{k'\ne j,l}u^*_{k'(jl)}v^*_{k'(jl)}G_{kk'} \\
-\sum_{k\ne j,l}\mid v_{k(jl)}\mid^{4}G_{kk}+
\epsilon_{j}-sN_{i_j}\lambda+\epsilon_{l}-sN_{i_l}\lambda\\
=E_{jl}-2\sum_{k\ne j,l} \rho_{k(jl)}sN_{i_k}\lambda-sN_{i_j}\lambda-sN_{i_l}\lambda,
\end{array}
\end{equation}
and $T_\gamma$ are state dependent energy terms 
\begin{equation}
\begin{array}{c}
 T_{k(0)}={i\hbar\over 2}(v_{k(0)}^{*}\dot{v}_{k(0)}-\dot{v}_{k(0)}^{*}v_{k(0)})\\
=2\mid v_{k(0)}\mid^2(\epsilon_{k}-sN_{i_k}\lambda)-2G_{kk}\mid v_{k(0)}\mid^{4}\\
+{\Delta_{k(0)}^{*}\over 2}\left({\mid v_{k(0)}\mid^{4}\over (v_{k(0)}u_{k(0)})^{*}}-v_{k(0)}u_{k(0)}\right)+
{\Delta_{k(0)}\over 2}\left({\mid v_{k(0)}\mid^{4}\over v_{k(0)}u_{k(0)}}-(v_{k(0)}u_{k(0)})^{*}
\right);
\end{array}
\end{equation}

\begin{equation}
\begin{array}{c}
 T_{k(jl)}={i\hbar\over 2}(v_{k(jl)}^{*}\dot{v}_{k(jl)}-\dot{v}_{k(jl)}^{*}v_{k(jl)})\\
=2\mid v_{k(jl)}\mid^2(\epsilon_{k}-sN_{i_k}\lambda)-2G_{kk}\mid v_{k(jl)}\mid^{4}\\
+{\Delta_{k(jl)}^{*}\over 2}\left({\mid v_{k(jl)}\mid^{4}\over (v_{k(jl)}u_{k(jl)})^{*}}-v_{k(jl)}u_{k(jl)}\right)+
{\Delta_{k(jl)}\over 2}\left({\mid v_{k(jl)}\mid^{4}\over v_{k(jl)}u_{k(jl)}}-(v_{k(jl)}u_{k(jl)})^{*}
\right),
\end{array}
\end{equation}
where the notation $\Delta_\gamma$ defined in Rel. (\ref{notatii}) was introduced.

The time dependent equations are obtained by minimizing the
functional (\ref{expr0}), that is,  differentiating with respect
the independent parameters $v_{k(0)}$, $v_{k(ij)}$,
$c_{0}$ and $c_{jl}$, together with their complex conjugates.
The Eqs. (\ref{ec1})-(\ref{ec2}) were derived in different
ways in Refs. \cite{koonin,blocki,prc08}, and it is straightforward
to introduce the condition (\ref{coop}). So, we will focus on the derivation
of Eqs. (\ref{ec3})-(\ref{ec6}), related to the mixing of configurations.

To obtain the derivatives, the following relations must be used:
\begin{equation}
\begin{array}{c}
\mid c_{0}\mid^{2}+\sum_{j,l\ne j} \mid c_{jl}\mid^{2}=1;\\
{\partial\over \partial c_{0}}
\left[\mid c_{0}\mid^{2}+\sum_{j,l\ne j} \mid c_{jl}\mid^{2}\right]=0;\\
\dot{c}_{0}c_{0}^{*}+\sum_{j,l\ne j}\dot{c}_{jl}c_{jl}^{*}=
-\dot{c}_{0}^{*}c_{0}+\sum_{j,l\ne j}\dot{c}_{jl}^{*}c_{jl};\\
{\partial \dot{c}_{0}c_{0}^{*}\over \partial c_{0}}=-\dot{c}_{0}^{*};\\
{\partial \dot{c}_{jl}c_{jl}^{*}\over \partial c_{jl}}=-\dot{c}_{jl}^{*}.
\end{array}
\end{equation}
The previous conditions ensure that the sum of the probabilities $\mid c_\gamma\mid^2$
of all configurations
is one.

The first equations are obtained by differentiating with respect $c_0$ and
$c_0^*$

\begin{equation}
\begin{array}{c}
\label{dife1}
c_{0}^{*}\bar{E}_{0}+i\hbar\dot{c}_{0}^{*}-c_{0}^{*}\sum_{k}T_{k(0)}+
\sum_{l,j\ne l}h_{jl}c_{jl}^{*}=0;
\end{array}
\end{equation}

\begin{equation}
\begin{array}{c}
\label{dife2}
c_{0}\bar{E}_{0}-i\hbar\dot{c}_{0}-c_{0}\sum_{k}T_{k(0)}+
\sum_{l,j\ne l}h_{jl}c_{jl}=0.
\end{array}
\end{equation}

After multiplying by $c_0$ the Rel. (\ref{dife1}) and by $c_0^*$  the Rel. (\ref{dife2})
and subtracting we obtain:

\begin{equation}
\begin{array}{c}
\label{em6}
i\hbar[c_{0}\dot{c}_{0}^{*}+c_{0}^{*}\dot{c}_{0}]=
\sum_{l,j\ne l}h_{jl}[c_{jl}c_{0}^{*}-c_{jl}^{*}c_{0}]
\end{array}
\end{equation}

Similar equations are obtained by differentiating the functional
with respect $c_{jl}$ and $c_{jl}^{*}$

\begin{equation}
\begin{array}{c}
\label{difx}
c_{jl}^{*}\bar{E}_{jl}+i\hbar\dot{c}_{jl}^{*}-c_{jl}^{*}\sum_{k\ne j,l}T_{k(jl)}+
h_{jl}c_{0}^{*}=0;
\end{array}
\end{equation}

\begin{equation}
\begin{array}{c}
\label{difxx}
c_{jl}\bar{E}_{jl}-i\hbar\dot{c}_{jl}-c_{jl}\sum_{k\ne j,l}T_{k(jl)}+
h_{jl}c_{0}=0.
\end{array}
\end{equation}

After multiplying by $c_{jl}$ the Rel. (\ref{difx}) and by $c_{jl}^*$ 
the Rel. (\ref{difxx}) and subtracting we construct the expression:

\begin{equation}
\begin{array}{c}
\label{em7}
i\hbar[c_{jl}\dot{c}_{jl}^{*}+c_{jl}^{*}\dot{c}_{jl}]=
h_{jl}[c_{jl}^{*}c_{0}-c_{jl}c_{0}^{*}].
\end{array}
\end{equation}

In a similar way we determine an exchange term between seniority zero
and seniority one configurations. We start from the equalities
obtained from the identities (\ref{dife1}) and (\ref{difx})
\begin{equation}
\begin{array}{c}
i\hbar \dot{c}_{0}c_{jl}^{*}=c_{0}c_{jl}^{*}\bar{E}_{0}- c_{0}c_{jl}^{*}
\sum_{k}T_{k(0)}+\sum_{m,n\ne m}h_{mn}c_{mn}c_{jl}^{*};\\
i\hbar \dot{c}_{jl}^{*}c_{0}=-c_{0}c_{jl}^{*}\bar{E}_{jl}+ c_{0}c_{jl}^{*}
\sum_{k\ne j,l}T_{k(jl)}-h_{jl}c_{0}c_{0}^{*};\\
i\hbar \dot{c}_{0}^{*}c_{jl}=-c_{0}^{*}c_{jl}\bar{E}_{0}+ c_{0}^{*}c_{jl}
\sum_{k}T_{k(0)}-\sum_{m,n\ne l}h_{mn}c_{jl}c_{mn}^{*};\\
i\hbar \dot{c}_{jl}c_{0}^{*}=c_{0}^{*}c_{jl}\bar{E}_{jl}- c_{0}^{*}c_{jl}
\sum_{k\ne j,l}T_{k(jl)}+h_{jl}c_{0}c_{0}^{*},
\end{array}
\end{equation}

and we construct the expression

\begin{equation}
\begin{array}{c}
\label{em8}
i\hbar{d (c_{jl}^{*}c_{0})\over dt}=c_{0}c_{jl}^{*}(\bar{E}_{0}-\bar{E}_{jl})
+ c_{0}c_{jl}^{*}\left(\sum_{k\ne j,l}T_{k(jl)}-\sum_{k}T_{k(0)}\right)\\
+\sum_{\{mn\}\ne\{jl\}} h_{mn}c_{mn}c_{jl}^{*}+
h_{jl}(c_{jl}c_{jl}^{*}-c_{0}c_{0}^{*}).
\end{array}
\end{equation}

Another exchange term is produced between  seniority two states:

\begin{equation}
\begin{array}{l}
-i\hbar\dot{c}_{jl}^{*}=c_{jl}^{*}\bar{E}_{jl}+c_{jl}^{*}\sum_{k\ne j,l}T_{k(jl)}
+h_{jl}c_{0}^{*},\\
i\hbar\dot{c}_{mn}=c_{mn}\bar{E}_{mn}+c_{mn}\sum_{k\ne m,n}T_{k(mn)}
+h_{mn}c_{0},
\end{array}
\end{equation}
giving eventually the next relation 
\begin{equation}
\begin{array}{c}
\label{em9}
i\hbar(c_{jl}^{*}\dot{c}_{mn}+\dot{c}_{jl}^{*}c_{mn})=
c_{mn}c_{jl}^{*}(\bar{E}_{mn}-\bar{E}_{jl})\\
+c_{mn}c_{jl}^{*}
\left(\sum_{k\ne m,n}T_{k(mn)}-\sum_{k\ne j,l}T_{k(jl)}\right)+h_{mn}c_{0}c_{jl}^{*}
-h_{jl}c_{0}^{*}c_{mn}.
\end{array}
\end{equation}

If the notations (\ref{notatii}) are used in the
expression (\ref{em6}), (\ref{em7}), (\ref{em8}) and (\ref{em9}), the time  dependent equations (\ref{ec3})-(\ref{ec6})
for the dynamical pair breaking effect with dynamic projection of number of particles are obtained.








\end{document}